\documentclass[pss,fleqn,embeddedheads]{w-art}
\usepackage{times}
\usepackage{w-thm}
\usepackage[]{graphicx}
\begin{document}
\DOIsuffix{theDOIsuffix}
\Volume{XX}
\Issue{1}
\Copyrightissue{01}
\Month{01}
\Year{2004}
\pagespan{1}{}



\title{Atomic vibrations in disordered systems: Comparison of disordered diamond lattices and a realistic amorphous silicon model}


\author{J. K. Christie\footnote{Corresponding
     author: e-mail: {\sf jkc25@cam.ac.uk}}}
\address{Department of Chemistry, University of Cambridge, Lensfield Road, 
Cambridge, CB2 1EW, UK}
\author{S. N. Taraskin}
\author{S. R. Elliott}

\begin{abstract}
Force-constant and positional disorder have been introduced 
into diamond lattice models in an attempt to mimic the 
vibrational properties of a realistic amorphous
silicon model. Neither type of disorder is sufficient on its own to mimic
the realistic model.  By comparing the spectral densities of these models,
it is shown that a combination of both disorders is a better
representation, but still not completely satisfactory.  Topological 
disorder in these
models was investigated by renumbering the atoms and 
examining the dynamical matrix graphically.  The dynamical matrix of
the realistic model is similar to that of a positionally-disordered
lattice model, implying that the short-range order in both systems is similar.
\end{abstract}
\maketitle                   
\renewcommand{\leftmark}
{J. K. Christie, S. N. Taraskin and S. R. Elliott: Atomic vibrations 
in disordered systems etc.}

\section{Introduction}

The physics of atomic vibrations in perfect crystals is well understood 
\cite{Maradudin et al, Dove}.
In the presence of disorder (e.g. noncrystallinity), many extra phenomena 
are observed: for
example, localization \cite{John et al}, the boson peak (extra modes at 
low frequencies) \cite{Taraskin et al},
and anomalies in the specific heat \cite{Feldman et al}
and thermal conductivity \cite{Pohl et al}. The microscopic origin of 
these and other, similar, phenomena have been difficult to deduce. 
In particular, it 
would be helpful to know the contributions due to different types of disorder.
In this paper, we consider the
effects of three types of disorder: {\it force-constant disorder},
in which the strengths of the springs between the atoms are disordered,
{\it positional disorder}, in which the atomic positions are
disordered, and {\it topological disorder}, in which the network of
bonds is disordered. We compare the behaviour of models with force-constant
and positional 
disorders (separately and in combination) to those of a realistic atomic 
model. A further aim is to see if it is possible to mimic the behaviour 
of amorphous systems with an appropriately-disordered crystal.

\section{Force-constant and positional disorder}

In this paper, we study 3D diamond-cubic crystal lattices, in which all atoms
are connected by springs, and have the same mass.  
We assume that the displacements
from equilibrium are small, and use the {\it harmonic approximation}.
Hence, we can treat the system classically \cite{Maradudin et al}, and
describe the vibrations by a Hermitian dynamical matrix $\hat{\bf{D}}$.
The problem can then be treated in the Hamiltonian formalism
\cite{Elliott et al, Ehrenreich and Schwarts, Economou}, with
energy $\epsilon=\omega^{2}$, the squared vibrational frequency.
We study the effect of disorder on the {\it spectral density}:
$\hat{{\bf A}}(\epsilon)=\left\langle\delta(\epsilon-\hat{{\bf D}})
\right\rangle$,
where $\left\langle\ldots\right\rangle$ represents configurational averaging. 
We can represent the spectral-density operator in the crystalline basis 
$|\bf{k},\beta\rangle$,
where $\bf{k}$ is the wavevector of a plane wave, and $\beta$ is the
polarization \cite{Taraskin and Elliott}. In this case, the spectral
density $A_{\bf{k}\beta}(\epsilon)$ is:
\begin{equation}
A_{\bf{k}\beta}(\epsilon)=\left\langle\sum_{d}|\left\langle d|\bf{k},
\beta\right\rangle|^{2}\delta(\epsilon-\epsilon_{d})\right\rangle,
\end{equation}
where $|\left\langle d|\bf{k},\beta\right\rangle|^{2}$ are the weights
of the (disordered) eigenstate $|d\rangle$ (with energy $\epsilon_{d}$)
in the crystalline eigenstate $|\bf{k},\beta\rangle$. 
Large ($\sim 10^{5}$ atoms) diamond-cubic crystals have been simulated, with
nearest-neighbour spring constant $\kappa_{1}$ and next-nearest-neighbour 
spring constant $\kappa_{2}$.  A ratio of $\kappa_{1}/\kappa_{2}=8$ was 
chosen, as this
gave a vibrational density of states (VDOS) most like that of a crystal 
with the
more realistic modified Stillinger-Weber (mSW) potential \cite {SW,mSW}.
Force-constant disorder was added by randomly taking
the spring constants from a uniform distribution of half width 
$\Delta$, centred
on the crystalline spring-constant value. The width $\Delta$ is expressed as 
a fraction of
the crystalline spring-constant value $\kappa$, and this fraction is the 
same for both 
spring constants. To avoid mechanical instabilities, widths were limited 
such that all force constants were positive, i.e. $\Delta\leq\kappa$.
Positional disorder was introduced by randomly
shifting all atoms by distances, chosen from Gaussian distributions centred
on zero with half width $\sigma$, independently in the $x$-, $y$- 
and $z$-directions.

Figure \ref{fig:1} shows the dispersion relation for a
positionally-disordered diamond lattice with $\sigma=0.1r_{1}$, where $r_{1}$
is the nearest-neighbour distance.  At each
value of $k$, the spectral density has been orientationally averaged, and 
the positions of the peaks are shown in the figure. The
dispersion relation shows many of the expected features: three acoustic
branches are observed, of which one is longitudinal, and two are 
transverse. Mixing between 
the polarisations is seen: the longitudinal spectral density
picks up the peaks due to the transverse acoustic (TA) branches.  The
optic modes are less clearly defined; the peaks due to 
these modes were broadened considerably by even small amounts of disorder.
Nevertheless, the presence of optic branches at high energy in both 
the transverse and longitudinal spectral densities is observed.

\begin{vchfigure}
\includegraphics[height=3.25in,angle=270]{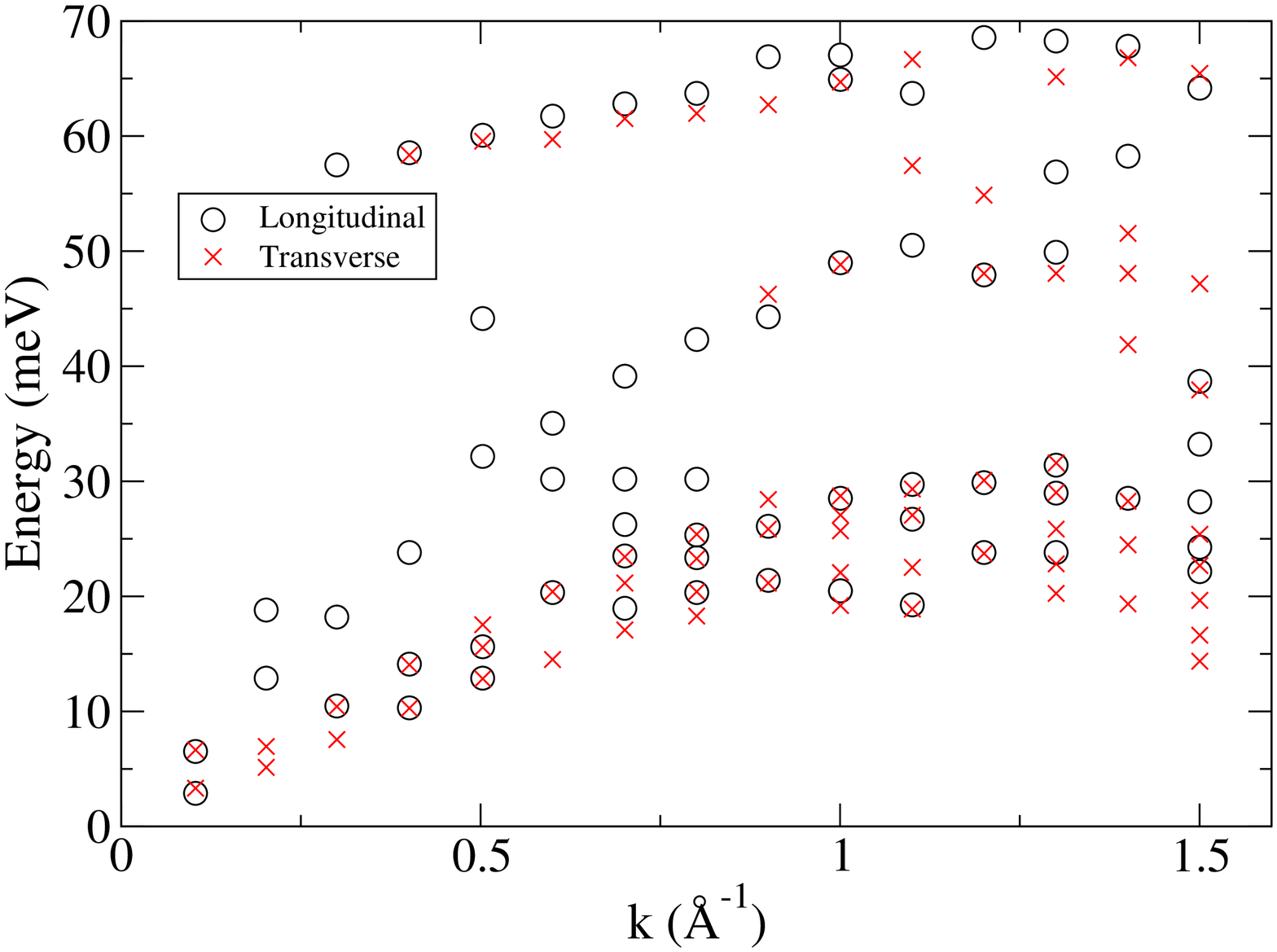}
\vchcaption{Longitudinal (circles) and transverse (crosses) dispersion 
relation for a positionally-disordered diamond lattice
with $\sigma = 0.1r_{1}$, where $r_{1}$ is the nearest-neighbour distance.
 Mixing between polarisations is clear.}
\label{fig:1}
\end{vchfigure}

The effect of a low degree of positional disorder on the spectral density is 
qualitatively very similar to the effect of small force-constant 
disorder \cite{TLNE}; the peak shifts in position, and broadens. 
When compared to the spectral density of a realistic 4096-atom model 
of amorphous silicon, generated with a modified WWW procedure 
\cite{WWW,Barkema and Mousseau} and relaxed under the mSW potential, it
was observed 
that neither type of disorder was adequate on its own to reproduce the 
features of the amorphous model's spectral density.

A combination of these two types of disorder is better able to reproduce 
the spectral density, as seen in Figure \ref{fig:2}.  The disordered
lattice models are able to mimic the main features of the amorphous model,
with peak positions and shapes at least roughly correct, but
they often have more peaks (particularly at high $k$) than the rather smoother 
amorphous spectral densities. This is due to the orientational 
averaging - each
of these peaks comes from a different orientation of the $\bf{k}$-vector.
The fact that these peaks are at different energies suggests that, despite
the added disorder, the lattice models are still anisotropic,
whereas the amorphous model is isotropic (since the multiple
peaks are absent). This implies that a form of disorder needs to be
considered that will render the lattice models less anisotropic. 

\begin{vchfigure}
\includegraphics[width=4.0in,angle=270]{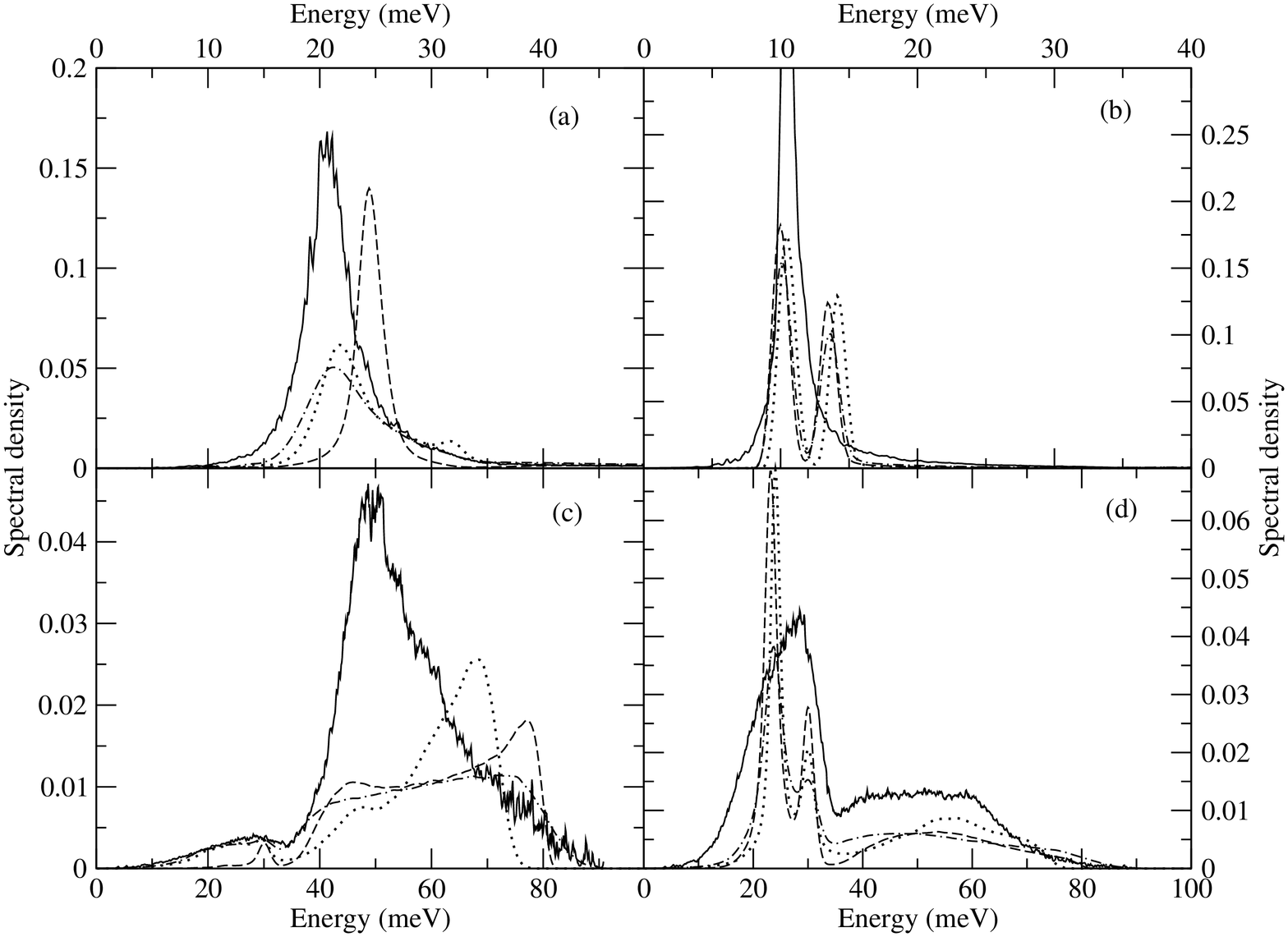}
\vchcaption{Orientationally-averaged spectral densities of a realistic 
4096-atom amorphous silicon model compared to disordered 125 000-site diamond 
lattices. Each graph shows the spectral densities of the realistic model 
(solid line), a lattice model 
with force-constant disorder $\Delta=0.75\kappa$ (dashed line), a lattice
model with positional disorder $\sigma=0.15r_{1}$ (dotted line), and a model
with both the above force-constant and positional disorders (dot-dashed line). 
$k=0.401$\AA$^{-1}$ ($k=0.411$\AA$^{-1}$ for the realistic model):
(a) longitudinal, (b) transverse; $k=1.200$\AA$^{-1}$ ($k=1.197$\AA$^{-1}$
for the realistic model): (c) longitudinal, (d) transverse.  The value of $k$
is slightly different between the lattice and realistic models due to the
necessity of using $\mathbf{k}$-vectors that fit the periodic boundary
conditions.}
\label{fig:2}
\end{vchfigure}

\section{Topological disorder}

Another type of disorder is topological disorder, in which the bond network is
disordered such that the ring-size distribution is no longer crystalline. 
Topological disorder was examined by studying the dynamical matrix.
If two atoms $i$ and $j$ interact, then the dynamical matrix elements 
representing that interaction will be non-zero. We make use
of a simple atom renumbering scheme and show the dynamical matrix graphically,
to compare the topological disorder present in the realistic model and in
positionally-disordered models.

Atoms were renumbered by first dividing the models into small cubes, 
such that the average occupation was 
one atom per small cube.  The models are cubic, with side $L$.
The origin was taken at one corner of the model (say $x,y,z=0$), and atoms were
numbered by moving up through all the small cubes above it in
the positive $z$-direction, numbering the atoms in order of the $z$-coordinate
of their position. Once the $z=L$ face of the model was reached, 
the scheme returned to the $z=0$ face,
and moved along one small cube distance in the $y$-direction. As before, 
atoms were numbered by moving up in the $z$-direction.  This process 
was repeated until the $y=L$ face was reached, when the scheme
returned to $y,z=0$ and shifted along one small cube in the $x$-direction,
and then repeated the numbering above, until all atoms were numbered. 
The atom number now corresponds (in some way) to its position.

\begin{vchfigure}
\begin{tabular}{ccc}
(a) \includegraphics[width=1.5in]{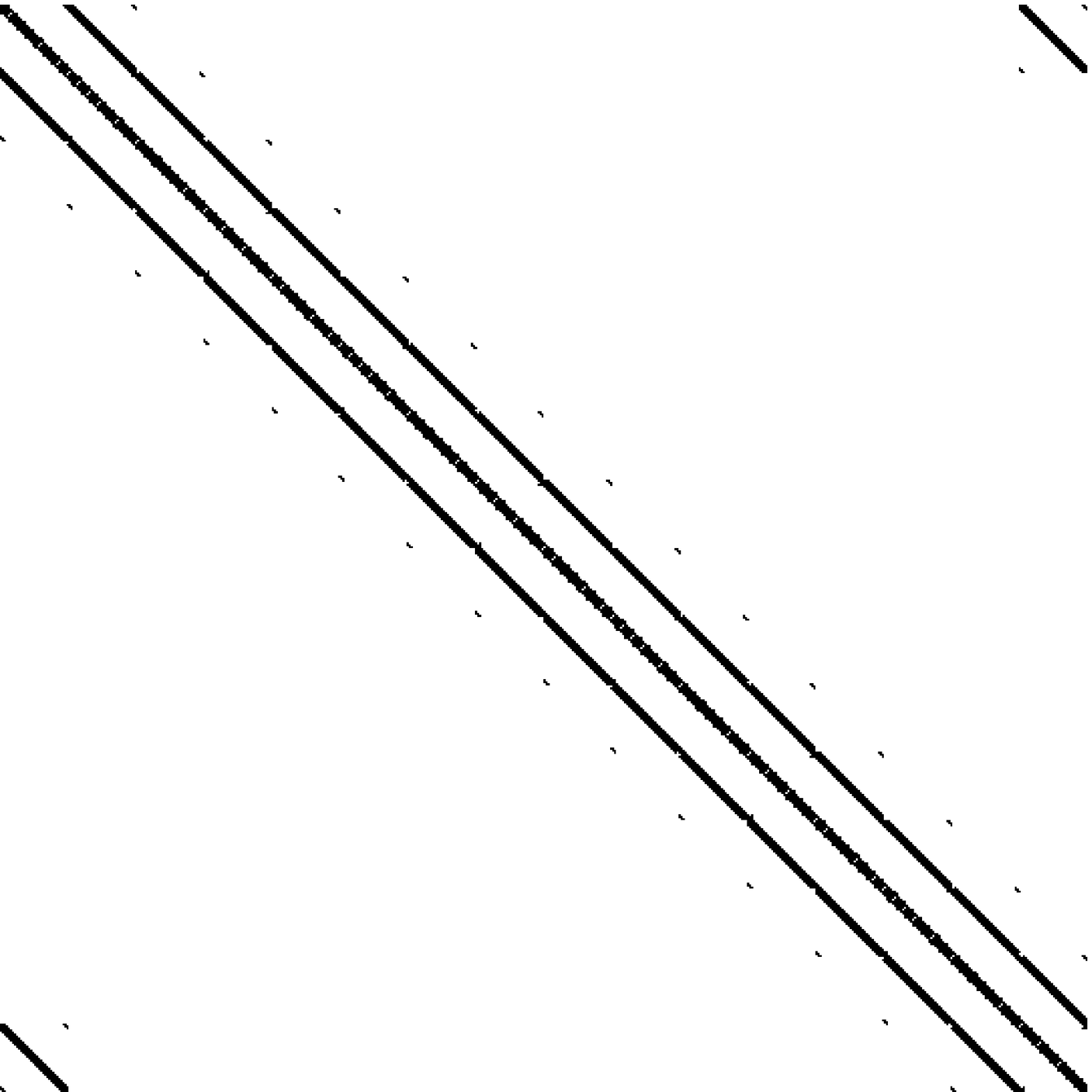}&
(b) \includegraphics[width=1.5in]{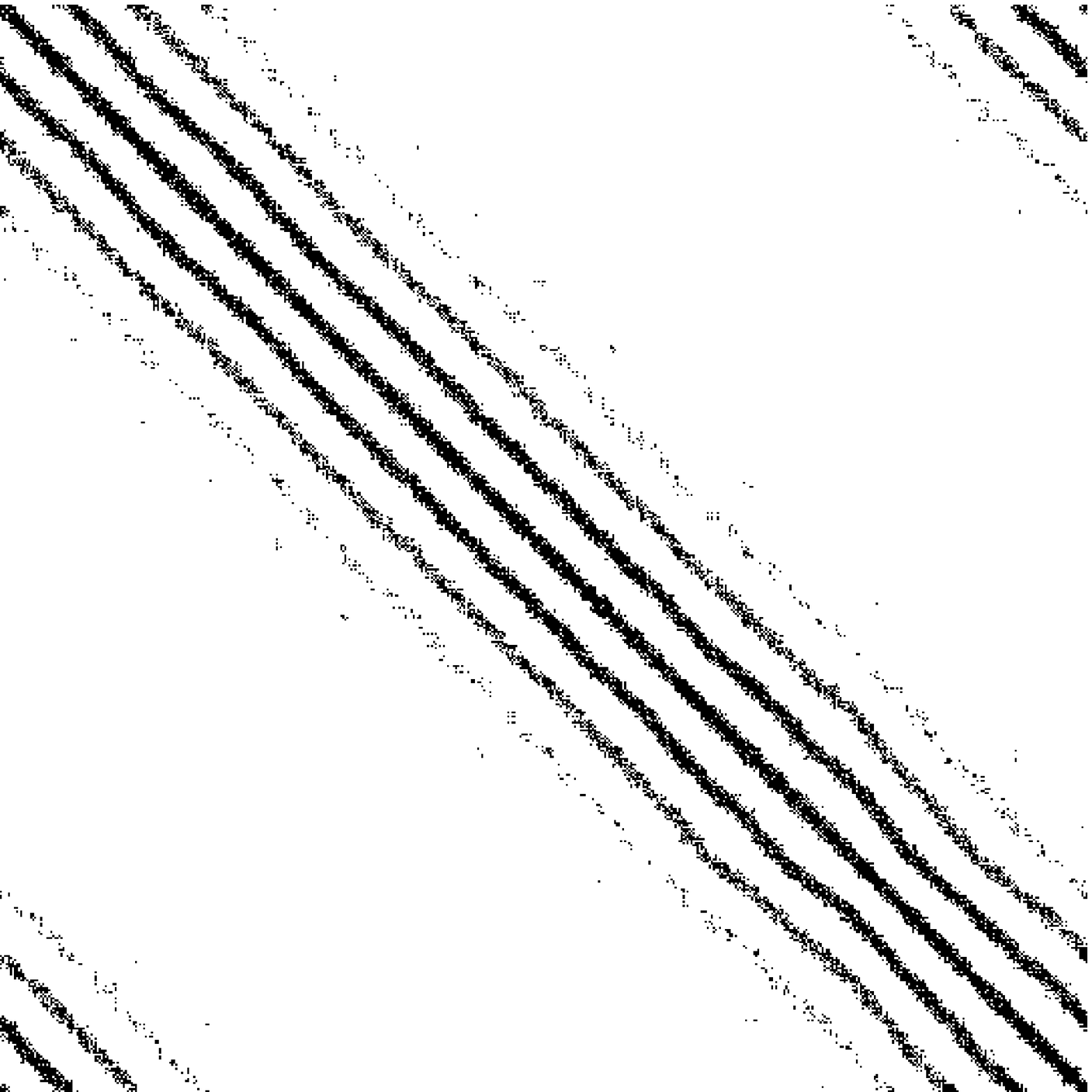}&
(c) \includegraphics[width=1.5in]{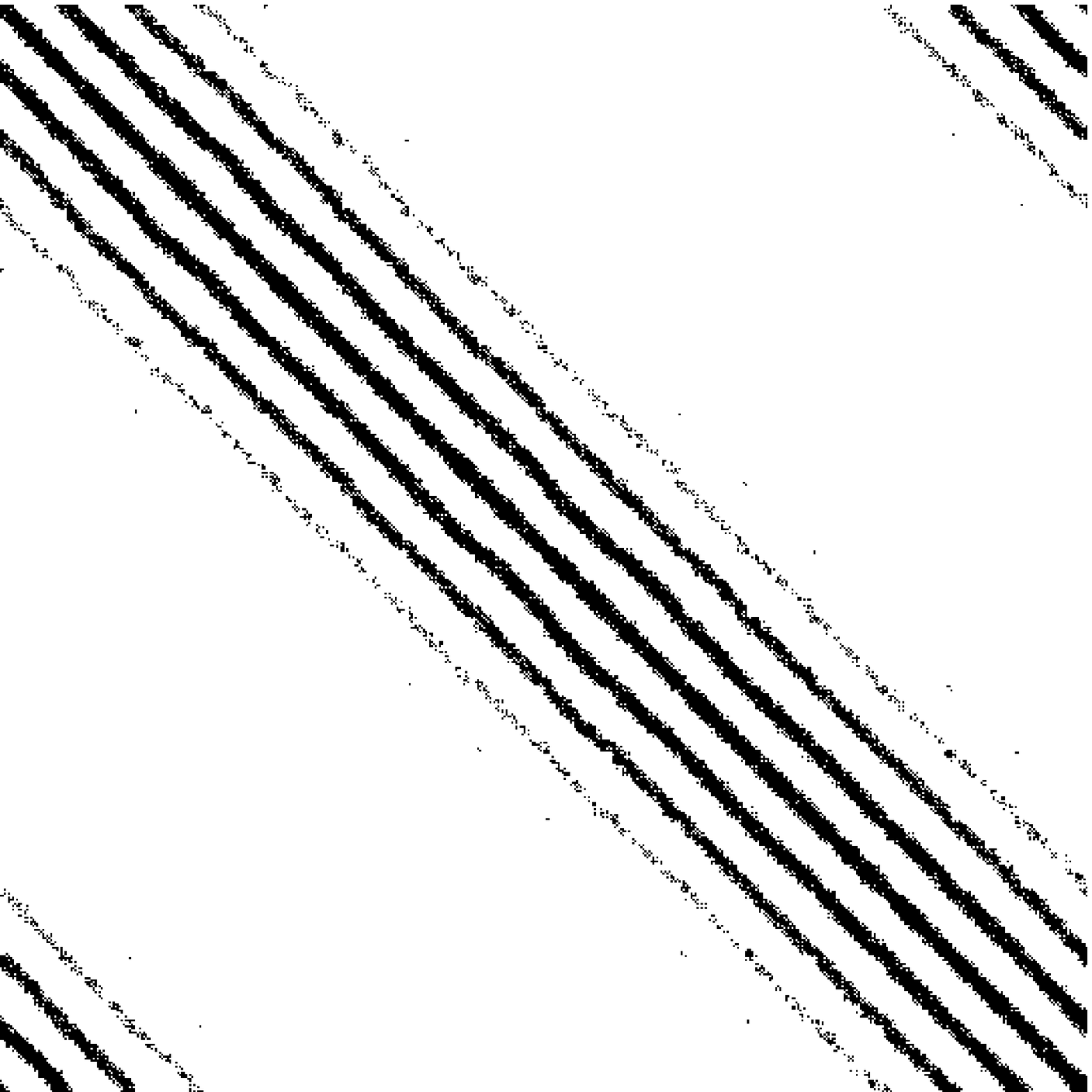}\tabularnewline
\end{tabular}
\vchcaption{Graphical representation of the dynamical matrix of:
(a) a diamond crystal with nearest- and next-nearest-neighbour interactions;
(b) the same crystal positionally disordered with $\sigma=0.6r_{1}$;
(c) a realistic 4096-atom model of amorphous silicon.} \label{fig:3}
\end{vchfigure}

In this renumbering scheme, the dynamical matrix of a crystalline diamond 
lattice with nearest-
and next-nearest neighbour interactions is shown in Fig.\ \ref{fig:3}a.
The effect of introducing positional disorder ($\sigma=0.6r_{1}$),
into the diamond lattice, is shown in Fig.\ \ref{fig:3}b.  The realistic model 
of amorphous silicon, with the same renumbering scheme,
is shown in Fig.\ \ref{fig:3}c.  Remarkably, the dynamical matrix for the
realistic model (Fig.\ \ref{fig:3}c) looks very similar to that of the
positionally-disordered crystal
(Fig.\ \ref{fig:3}b).  We interpret this as implying that the 
short-range structural order, responsible for the vibrational behaviour, 
in the topologically-disordered amorphous model is very close to that in a 
positionally-disordered crystal.

\section{Conclusion}
Examining the spectral densities of diamond lattice models with added
force-constant and positional disorders shows that these disorders are
insufficient (either by themselves or in combination) to mimic completely the
vibrational 
behaviour of a realistic model of amorphous silicon.  Representing the
neighbour list graphically shows that the short-range order, responsible for
the vibrational behaviour, in a
positionally-disordered diamond lattice is similar to that of a realistic
model, and this is an avenue for further research.

\begin{acknowledgement}
JKC is grateful to the Engineering and Physical Sciences Research
Council for financial support. We thank G.\ T.\ Barkema
for providing us with the atomic co-ordinates of the amorphous model 
used.
\end{acknowledgement}

\end{document}